\documentstyle[aps,prb]{revtex}
\tightenlines
\begin{document}  
\draft

\title{Transport properties of the hierarchical model for 
stretched polymers}
\author{Chen-Ping Zhu$^{1,2}$, Shi-Jie Xiong$^{1,*}$ and Tao Chen$^1$}
\address{ $^1$ National Laboratory of Solid State Microstructures 
and Department of Physics, 
Nanjing University, 
Nanjing 210093, People's Republic of China \\
$^2$ College of Science, Nanjing University of 
Aeronautics and Astronautics,
Nanjing 210016, People's Republic of China.}
\maketitle

\begin{abstract}
    We prove that the hierarchical fractal model recently proposed for 
describing the stretched polymers [A. N. Samukhin {\it et al},
Phys. Rev. Lett.
{\bf 78}, 326(1997)] is 
equivalent to a one-dimensional chain with hierarchical aperiodic 
structure. By use of the transfer matrix technique we calculate
the electronic transmission and the dc conductance. 
We find that there exist sharp-edged transmission subbands and gaps, but 
the transmission subbands are rich in substructures which show 
self-similarity. 
The temperature dependence of conductance 
$g (T)$ is sensitive to the variation of the Fermi level and to the 
structure parameters $m$ and $n$ of the original 
fractal structure. The relationship of the obtained results to 
the experimental data and other theoretical works is discussed. 
\end{abstract}
\pacs{ PACS numbers: 73.61.Ph,05.60.+w,71.30.+h,72.90.+y \\ 
 $^*$E-mail: sjxiong@nju.edu.cn}

\section{Introduction}
    It is widely accepted that polymers represent complicated structures
which consist of coupled one-dimensional (1D) chains, and many unusual
properties can be understood based on their quasi-1D feature. During last
decades analytical and numerical studies revealed that the phenomenon of
Anderson localization is a characteristic feature of 1D systems. For a
quasi-1D system the interchain coupling tends to delocalize the electronic
states, which may lead to extended states when interchain diffusion rate
$\omega$ exceeds a threshold value $\omega_{c}$\cite{s1}. 
Moreover, conductance 
of several polymers, such as stretched polyacetyllene and 
polyanilline, is enhanced
by a few orders of magnitude upon doping and can even reach high values
typical of metals, while the temperature dependence of static
conductance ($g_{dc}$) widely vary from sample to
sample\cite{s2,s3,s4,s5,s6,s7,s8}.

In order to understand the effect of chain windings and random 
interchain couplings on the electronic transport properties 
in quasi-1D polymers, 
Xiong {\it et al} \cite{s10,s100} have investigated the transmission 
coefficient and conductance for models of winding and randomly coupled 
polymers. It is proved that for such systems there exist conducting subbands 
with ``dilute'' structure which can account for the high values of 
conductance and its insulator-like temperature dependence of some 
polymers. 
On the other hand, Samukhin {\it et. al} \cite{s9} suggested that 
the peculiar features of conducting 
polymers can be derived based on a special hierarchical 
structure--a nearly 1D   
fractal. 

     In this paper we investigate the transmission 
spectrum of the hierarchical structure proposed in 
\cite{s9}. We prove that this fractal network is equivalent to a 
1D chain with hierarchical aperiodic 
structure for the transport of electrons in the stretched direction.
By solving the Schr{\"o}dinger equation of such an effective 1D  
model by the standard transfer matrix technique,
we obtain the 
transmission coefficient as a function of energy for 
different stages (or size) of the structure. We find that there also exist 
``dilute'' conducting subbands as those in Ref. \cite{s10} and \cite{s100}, 
but the structure in these subbands exhibits some kind of self-similarity 
and become more complicated when 
the stage of the system increases. As a result, the calculated conductance 
sensitively depends on 
the structure parameters $(m,n)$ of the original hierarchical fractal and 
on the position of the Fermi level.  

  The paper is organized as follows: In the next section we present 
the derivation of the equivalent 1D model and the basic formalism. 
In Section III the calculated transmission spectrum and conductance 
are illustrated. Section IV is devoted to a brief summary and discussion. 

\section{Basic formalism} 

     According to Samukhin {\it et. al}, 
the hierarchical model for the stretched polymer
is constructed in the following way: as a first stage, 
$n$ primitive bonds are taken to form an $n$-length chain 
along stretched direction, and 
$m$ threads of such chains are combined at both ends to 
form an $(m,n)$ boundle 
with two common terminals; in the following stages the boundles 
formed in the previous stage are taken as the elementary bonds and the 
same procedure is performed to form new boundles.
We are interested in the variation of the transmission spectrum in 
increasing the 
number of stages $N$, i.e. the size of 
the system. In measuring 
the conductance one needs to link electrodes to 
both ends, (see 
Fig.1). We use two integers, $(i,j)$, to label the sites, 
where $i$ indicates the position 
in the longitudinal direction and $j$ in the transverse direction. 
If the number of stages of the system is $N$, the 
longitudinal index $i$ is varying in the range of $[0,n^N]$. 
For a given $i$, the number of sites in the transverse direction is
\begin{equation}
\label{rho_N}
\rho_N(i)=\left\{ \begin {array}{l} 
m^N , \text{ if } \text{mod}(i,n)\neq 0 ;\\ 
m^{N-\alpha_N(i) } \text{ if } \text{mod}(i,n^{\alpha_N(i)})
=0 \text{ and } 
\text{mod}(i,n^{\alpha_N(i) +1})\neq 0, \alpha_N (i)
\text{ is integer in }(0,N); 
\\ 
1, \text{ if } i\leq 0 \text{ or } i\geq n^N, \end{array} 
\right. 
\end{equation}
where $\text{mod}(x,y)$ denotes the remainder on the division of $x$
by $y$. 
Thus, in the site index $(i,j)$, $j$ is ranging in $[1,\rho_N(i)]$. 
From the tight binding approach, 
the Schr\"{o}dinger equation can be written as
\begin{equation}
\label{ham}
E\psi_{i,j}= 
\epsilon_{i,j}\psi_{i,j}+\sum_{\delta (i,j)}t_{ij,\delta (i,j)}
\psi_{\delta (i,j)},
\end{equation}
where $\psi_{i,j}$ is the coefficient of the wave function on site 
$(i,j)$, $\delta (i,j)$ denotes the position of next neighbors of $(i,j)$, 
$\epsilon_{i,j}$ is its site energy, $t_{ij,\delta(i,j)}$ the hopping
between sites $(i,j)$ and $\delta (i,j)$. 
As we are only interested in the topological structure of the 
system, we can assume that the site energies and the nearest-neighbor 
hoppings are independent of the position. In this case 
one can set the site energy to be the energy origin and the hopping to be 
the energy unit. By summing over the transverse sites for a 
given $i$, Eq. (\ref{ham}) becomes 
\begin{equation}
\label{1d}
E\psi_i=\beta_N(i,i+1)\psi_{i+1}+\beta_N(i,i-1)\psi_{i-1},
\end{equation}
where 
\[
\psi_i=\frac{1}{\sqrt{\rho_N(i)}}\sum_{j=1}^{\rho_N(i)}\psi_{i,j},
\]
and 
\[
\beta_N(i,i')=\max \left( \sqrt{\frac{\rho_N(i)}{\rho_N(i')}},
\sqrt{\frac{\rho_N(i')}{\rho_N(i)}}\right).
\]
From Eq. (\ref{rho_N}) it is easy to see that 
\begin{equation}
\beta_N(i,i')=m^{|\alpha_N(i)-\alpha_N(i')|/2},
\end{equation}
where $\alpha_N(i)$ is the largest power of factor $n^{\alpha_N(i)}$ of integer
$i$ in the range $[1,n^N-1]$, 
as defined in Eq. (\ref{rho_N}), and we set $\alpha_N(i)=N$ for 
$i\leq 0$ and $i\geq n^N$. 

From Eq. (\ref{1d}) one can see that the fractal structure reduces to 
a 1D effective Hamiltonian with 
special aperiodic structure. In the derivation we 
have adopted the summation 
over the transverse sites which erases the degree of freedom of electrons
in this direction. This has no effect on the longitudinal transport 
properties for the one-channel incoming and outgoing leads of the 
present geometry. The situation becomes different if multichannels are 
connected to the system or there exists on-site or off-diagonal 
disorder in the system which can produce an energy-dependent terms 
in the reduced Hamiltonian. In this paper we only focus on the 
present geometry and the multichannel and disordered cases will 
be considered in separated ones. 

The transmission coefficient in such a one-channel structure 
can be calculated by the standard transfer matrix technique. 
The transfer matrix can be defined as follows 
\begin{eqnarray}
\left(\begin{array}{c}\psi_{i+1}\\ \psi_{i}\end{array}\right)= & 
\left(\begin{array}{cc}Em^{-\frac{|\alpha_N(i)-\alpha_N(i+1)|}{2}} & -m
^{\frac{|\alpha_N(i)-\alpha_N(i-1)|-|\alpha_N(i)-\alpha_N(i+1)|}{2}} 
\\ 1 & 0 \end{array}\right)
\left(\begin{array}{c}\psi_{i}\\ \psi_{i-1}\end{array}
\right) \nonumber \\   &  \equiv
B_N(i) \left(\begin{array}{c}\psi_{i}\\ \psi_{i-1}\end{array}\right).
\end{eqnarray} 
    So along the 1D chain with stage number $N$, 
the wave function of an electron in 
two leads are related to each other via a total transfer matrix 
\begin{equation}
\left(\begin{array}{c}\psi_{n^N+1}\\ \psi_{n^N}\end{array}\right)=
\prod_{i=0}^{n^N}B_N(n^N-i)\left(\begin{array}{c}
\psi_0\\ \psi_{-1}\end{array}\right)
\equiv {\bf T}(N) \left(\begin{array}{c}\psi_0\\ \psi_{-1}
\end{array}\right).
\end{equation}

       We demonstrate such an effective 1D chain structure in
Fig. 2. This is a one-to-one mapping from the original network 
of Fig.1, and the resultant 1D chain 
shares the hierarchical structure at its longitudinal direction 
and the mirror reflecting
symmetry as well. The self-similar features of the original 
structure are also kept in the mapping.

    The $\rho_N(i)$-structure in the longitudinal direction 
of the original system can be described 
by the following series:
\begin{equation}
1,\underbrace{m^N,\cdots,m^N}_{n-1},\underbrace{m^{N-1},
\underbrace{m^N,\cdots,m^N}_{n-1},\cdots,
m^N,\underbrace{m^{N-1},
\cdots,m^N}_{n-1}}_{n-1},m^{N-2},\cdots .
\end{equation}
One can see that this is a embedded multi-period structure. 
As a consequence the hoppings in the effective 1D chain are arranged as in 
the following series:
\begin{equation}
1,m^{\frac{N}{2}}\underbrace{1,\cdots,1}_{n-2},m^{\frac{1}{2}},
\underbrace{m^{\frac{1}{2}},
\underbrace{1,\cdots,1}_{n-2},m^{\frac{1}{2}},\cdots,
m^{\frac{1}{2}},
\underbrace{1,\cdots,1}_{n-2},m^{\frac{1}{2}}}_{n-2},
m^{\frac{1}{2}},
\underbrace{1,\cdots,1}_{n-2},m^{\frac{2}{2}},\cdots .
\end{equation}
Thus, one has the following recursion relation of the 
transfer matrix for $n\geq 2$:
\begin{eqnarray}
{\bf T}(N)= & B_N(n^N)B_N(n^N-1)B_{N-1}(n^{N-1}-1){\bf C}(N-1)
[{\bf C}(N-1)
\nonumber \\   & 
\times {\bf F}(N-1)]^{n-2}{\bf C}(N-1)B_{N-1}^{-1}(1)B_N(1)B_N(0),
\end{eqnarray}
where
\[
{\bf C}(N-1)=B_{N-1}^{-1}(n^{N-1}){\bf T}(N-1)B_{N-1}^{-1}(0),
\]
and 
\[
{\bf F}(N-1)=\left( \begin{array}{cc} \frac{E}{\sqrt{m^{N-1}}} & -1 \\ 1 & 0 
\end{array} \right) .
\]

      One can find that the 
effective 1D chain contains $n^N$ bonds among which there are 
$(n-2)n^{N-1}$ bonds of hopping 1, 2 bonds of hopping $\sqrt{m^N}$, 
and $2n^{N-i-1}(n-1)$ bonds of hoppings $\sqrt{m^i}$ for 
$i=1,2,\cdots,N-1$. These bonds are arranged in a special aperiodic 
way so that one may expect a self-similar and Cantor-set-like 
transmission spectrum. 
From the transfer matrix {\bf T}, we can easily calculate the 
transmission coefficient for electrons from the left 
to the right lead:
\begin{equation}
  \tau\left(E\right)=\frac{4\sin^2k}{[T_{21}-T_{12}+(T_{22}-T_{11})\cos k]^2+
(T_{11}+T_{22})^2 \sin^2k},    
\end{equation}
where $k=\arccos (E/2)$ is the wave vector of wave function in the leads. 
Thus, the conductance of the original network at the 
longitudinal direction is obtained according to Landauer formula
\begin{equation}
\label{lan}
  g(E)=\frac{e^2}{h}\int \tau(E) \frac {\partial f}{\partial E}dE
\end{equation}
where {\it f} is the Fermi distribution function
\begin{equation}
  f(E,T)=\frac{1}{1+\exp{(E-E_f)/k_BT}}. 
\end{equation}

\section{Transmission spectrum and temperature dependence of conductance}
      By using the above formulas we calculate the 
transmission coefficient as a function of energy for 
different structure parameters $(m,n)$ and 
different stage number.  
Typical results of transmission are illustrated in
Fig. 3 and Fig. 4. We can see that 
when the stage number of the structure is low (say
$N\leq 2$), $\tau (E)$ exhibits quite continuous variation with 
broad peaks and
dips. However, when the stage number $N\geq 3$ the whole energy interval
[-2,2] is divided into ``subbands'' which allow electron 
transmission and ``gaps'' of nearly zero transmission
with sharp edges. 
By increasing $N$, the subbands are further divided into ``minibands'' and 
``minigaps'', but the edges of the main subbands are almost 
independent of $N$. 
For large $N$, there are a large number of sparsely
distributed peaks with $\tau(E)$ equal to or near unity
which indicates almost complete transition corresponding to unscattered
electronic states. This structure of the transmission 
spectrum also shows the self-similarity and Cantor-set-like behavior, 
like the situation in the Fibbonaci series and other 
quasiperiodic or aperiodic systems.
By the comparison between Fig. 3 and Fig. 4, 
we find that although the number of the main subbands is the 
same but the width of the central main subband is slightly widened 
and the other two are shifted away from the band center 
by increasing $n$,  
which changes the number of the repeated bonds in the 
series. In the inset of Fig. 1(a) we also show the $N$ 
dependence of the transmission coefficient for energy 
$E=1.57$ which lies in the main band. It can be seen that
for this energy by increasing $N$ the transmission first decreases but then 
increases. This behavior is sensitive to the position of $E$ as can be seen 
from the complicated structure in the 
transmission spectrum for large $N$.

      For the polymer networks, an interesting problem is to 
investigate the scaling behavior and 
the temperature dependence of the dc conductance. 
Since the transmission subbands exhibit 
``dilute'' characteristics with a lot of transmission holes 
within them, the temperature dependence of the conductance 
is sensitive to the position of the Fermi level. From Eq. (\ref{lan}), 
for low temperature the conductance as a function of the Fermi
level shows almost the same characteristics as those for the 
transmission coefficient as a function of energy. From Figs. 3 and 
4 one can see that there exist energy points for which the transmission
of large system is greater than that of small system. As a consequence 
for the system with Fermi level at these points the conductance 
increases with the system size provided that the temperature is low 
enough. This may be regarded as the ``metallic'' scaling behavior. There 
are still many other energy points for which  the scaling behavior 
of the conductance is ``insulating'' as it decreases with the system size. 
However, the energy regimes for ``metallic'' and ``insulating'' states 
are strongly mixed with each other and form a complicated Cantor-set-like
structure. If the temperature becomes higher, such a structure is 
smoothened by the thermal averaging. In Figs. 5 and 6 we plot 
the Fermi level dependence of the conductance at temperature 
$T=400$K for different structures and 
different system size. It can be seen that at this temperature 
the curves become smooth and in the whole energy range the conductance 
decreases with the increase of the system size, showing the insulating 
behavior. 
This is because more transmission holes appear in the transmission 
subbands when $N$ increases and this reduces the average of the 
transmission. 
In Fig. 7 we plot the temperature dependence of 
conductance $g(T)$ for one structure but with different 
Fermi levels. 
The curves manifest that the conductance can increase or 
decrease with the decrease of the temperature, 
strongly depending on the Fermi level $E_f$. The  
orders of magnitude of the variation of conductance 
at the range of [50K, 400K] could vary from 1 to 4. Only 
for carefully chosen  values of $E_f$ (usually near edges of subbands for 
a given structure $(m,n))$, one could get a curve of $\log g(T)-T^{-1/2}$ 
whose slope can be fitted to the  
experimental data \cite{s1,s8}. In fact, the realistic stretched 
polymers should contain tremendous number of such networks 
in different stages, and doping might adjust Fermi level as well as the  
distribution of $(m,n)$. 
This may cause the deviation of the conductance calculated results in  
this simple model from the observed data. 
Nevertheless, as one can see from
Fig. 7, our calculation displays a curve of 
good fitting to expression  
$\ln g_{dc} \propto T^{-1/2}$ within quite large range of temperature, while
the solution of the model from the critical percolation 
approach in Ref. \cite{s8} 
demonstrates that $ g_{dc}\propto \frac{e^{-(T_0/T)^{1/2}}}{T^2}$ which 
slightly diverges from experimental results.

\section{Conclusions and Discussion}      

In summary, we have shown that the hierarchical fractal  
network of Fig. 1 is equivalent to a 1D aperiodic 
series if only one-channel transport properties are under consideration.
This equivalence is derived by summing over 
the transverse degree of freedom and, as a consequence, does not 
imply the one-dimensionality of this structure. 
The self-similarity and the aperiodicity of the original network are 
kept in this mapping. We calculate the transmission spectrum 
and temperature dependence of the conductance for systems with different
structure parameters and various size. The transmission 
spectrum manifests Cantor-set-like characteristics, with 
complicated structures made of mixed transmission peaks and gaps, like most 
of the 1D quasiperiodic and aperiodic systems. Such a structure 
is also reflected in the mixed 
insulating and metallic scaling behavior of the conductance 
at low temperature. At higher temperature such a structure in the 
conductance is smoothened by the thermal averaging and the 
scaling behavior is insulator like in the whole energy range. 
This implies some type of metal-insulator transition 
at special values of the Fermi level, in agreement with the analysis 
in Ref. \cite{s9}. 
The magnitude and temperature dependence of the conductance 
strongly depend on the position of the Fermi 
level.  By adjusting the value of $E_f$, one can 
reproduce the temperature dependence observed in the experiments. 
This simplified model can catch several features of polymer 
qualitatively but a complete description needs more sophisticated 
models.

\section*{Acknowledgments}
       This work is supported by National Fund of Natural Science of China.

\begin{figure}       
  \caption{ Hierarchical fractal network with structure 
   parameters $m=3$, $n=4$ and stage number (a) $N=1$, (b) $N=2$. }

  \caption{ The equivalent 1D chain for the hierarchical structure 
   shown in Fig. 1 with the same parameters. The numbers below the lines
   indicate the index of sites in the longitudinal direction, and 
   the numbers above the lines are the effective hopping integral of 
   the corresponding bonds. }

   \caption{ The transmission coefficient as a function of energy 
   for structures with $m=2$, $n=5$, and (a) $N=1$, (b) $N=2$, 
   (c) $N=3$, (d) $N=4$, (e) $N=5$. The inset of (a) shows 
   the $N$ dependence of $\tau (E)$ for $E=1.57$. }

   \caption{ The transmission coefficient as a function of energy 
   for structures with $m=2$, $n=6$, and (a) $N=1$, (b) $N=2$, 
   (c) $N=3$, (d) $N=4$. }

   \caption{ The conductance as a function of Fermi level at temperature 
   $T=400$K  
   for structures with $m=2$, $n=5$, and (a) $N=1$, (b) $N=2$, 
   (c) $N=3$, (d) $N=4$, (e) $N=5$. The hopping integral of the original 
   network is set to be 1eV. }

   \caption{ The conductance as a function of Fermi level at temperature 
   $T=400$K  
   for structures with $m=2$, $n=6$, and (a) $N=1$, (b) $N=2$, 
   (c) $N=3$, (d) $N=4$. The hopping integral of the original 
   network is set to be 1eV. }

   \caption{ The conductance as a function of temperature of system 
   with structure parameters $m=2$, $n=5$ and $N=5$ for different 
   Fermi levels. }

   \end{figure} 

\begin{references}
\bibitem{s1} Zhao H. Wang and Hamid H. S. Javadi, Phys. Rev. B {\bf 42},
5411 (1990).

\bibitem{s2} J. Joo, Oblakowski, and G. Du, Phys. Rev. B {\bf 49}, 2977 (1994).

\bibitem{s3} N. Basescu, Z -X.  Liu, D. Moses,  A. J. Heeger,  H. Narrmann,  and 
Theophilou,    Nature  (London) {\bf 327}, 403 (1987).

\bibitem{s4} J. Tsukamoto, A. Takahasi, and K. Kawasaki, Jpn. J. Appl. Phys. 
{\bf 29}, 125 (1990).

\bibitem{s5} Y. Nogami, H. Kaneko, T. Ishiguro, N. Hosoito, A. Takahashi, and 
J. Tsukamoto, Solid State Commun. {\bf 76}, 583 (1990).

\bibitem{s6} Y. Nogami, H. Kaneko, H. Ito, T. Ishiguro, T. Sasaki, N. Toyota,
A, Takahashi, and J. Tsukamoto, Phys. Rev. B {\bf 43}, 11829 (1991).


\bibitem{s7} H. H. Javadi, A. Chakraborty, C. Li, N. Theophilou, D. B. Swanson, A. G. MacDiarmid, 
and A. J. Epstein, Phys. Rev. B {\bf 43}, 2183 (1991).

\bibitem{s8} J. Joo, V. N. Prigodin, Y. G. Min, A. G. MacDiarmid and A. J. 
Epstein, Phys. Rev. B {\bf 50}, 12226 (1994).

\bibitem{s10} Shi-Jie Xiong and S. N. Evangelou, Phys. Rev. B {\bf 52}, 
R13079 (1995).

\bibitem{s100} Shi-Jie Xiong, Yan Chen, and S.N. Evangelou, Phys. Rev. Lett. 
{\bf 77}, 4414 (1996). 

\bibitem{s9} A. N. Samukhin, V. N. Prigodin, and L. Jastrabik, Phys. Rev. Lett.
{\bf 78}, 326 (1997).

\bibitem{s11} X -S. Chen and S -J. Xiong, Phys. Lett. A {\bf 179}, 217 (1993).

\end{references}
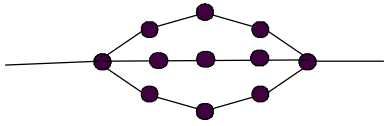
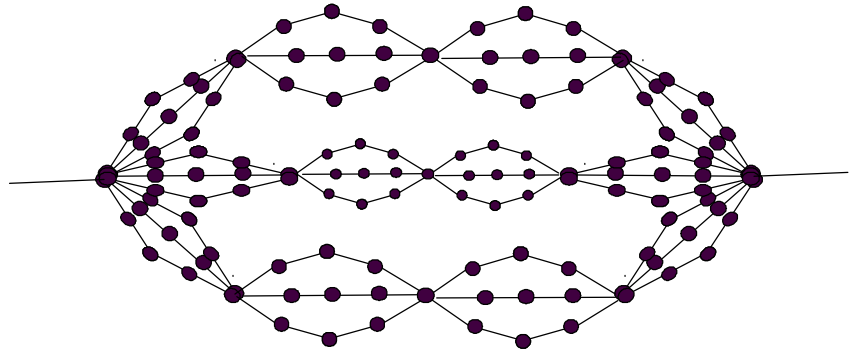
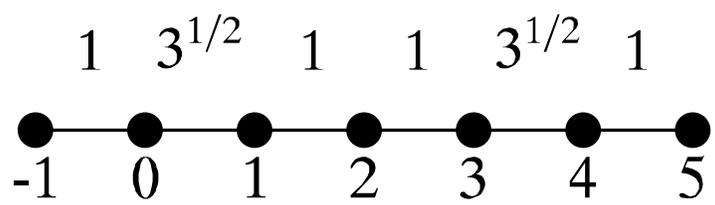
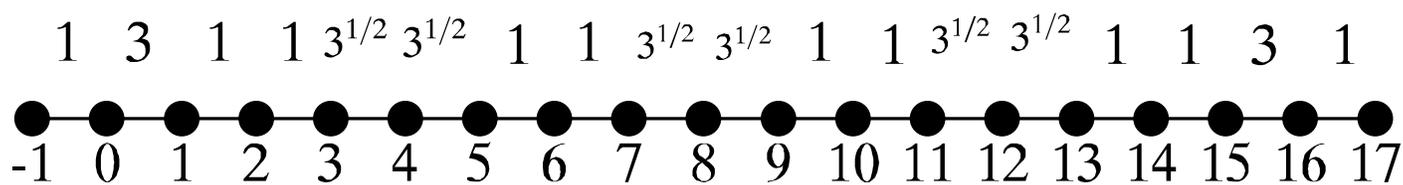
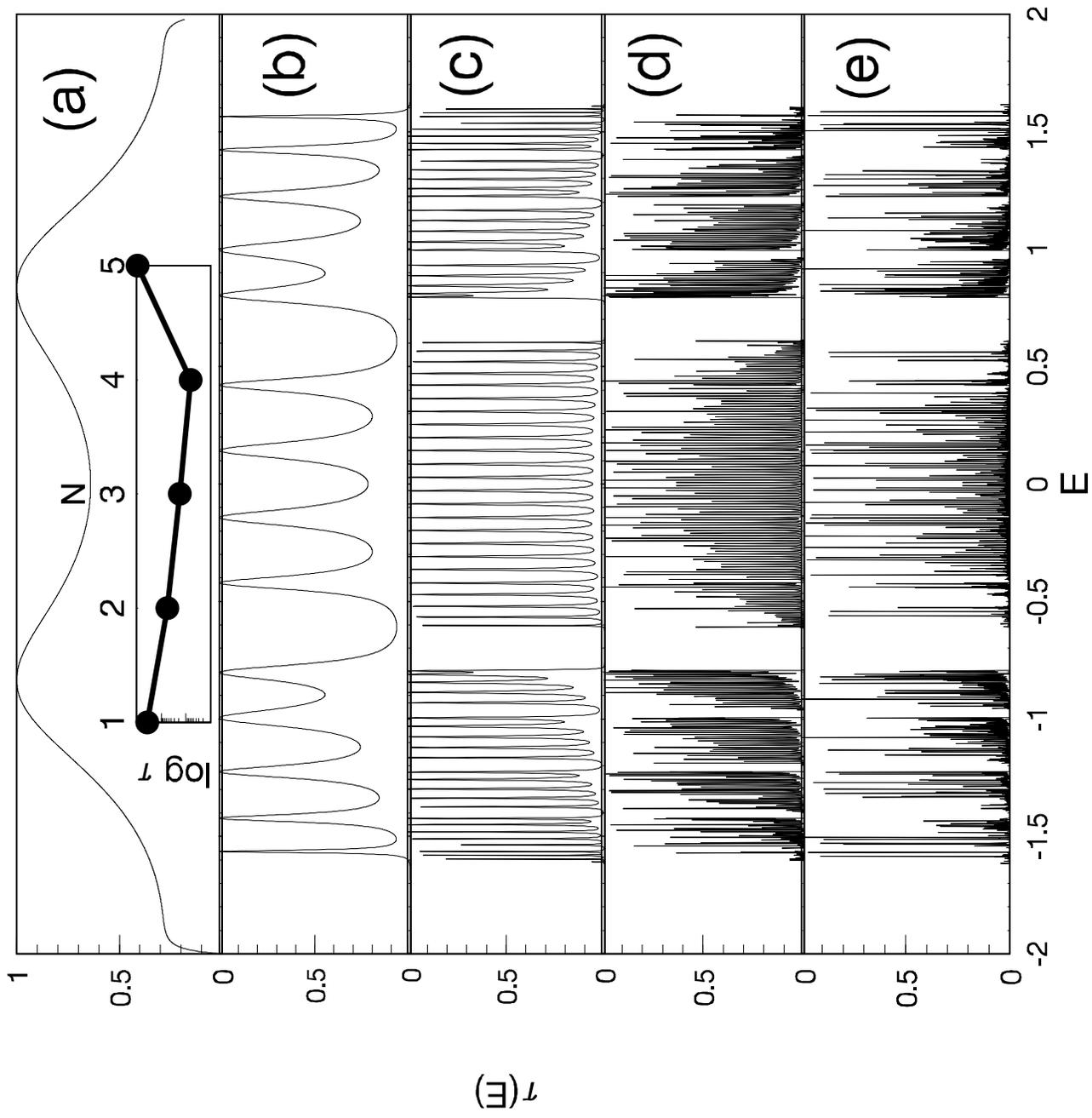
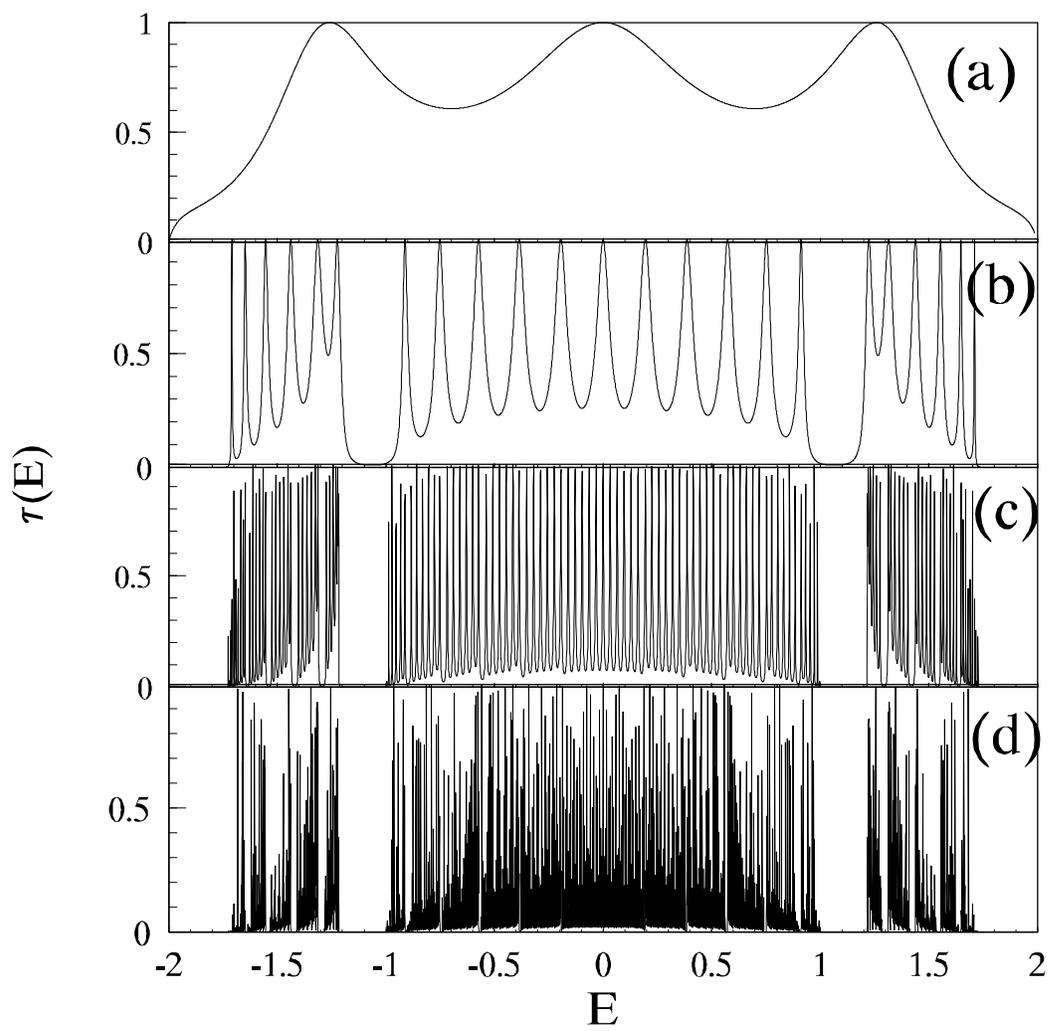
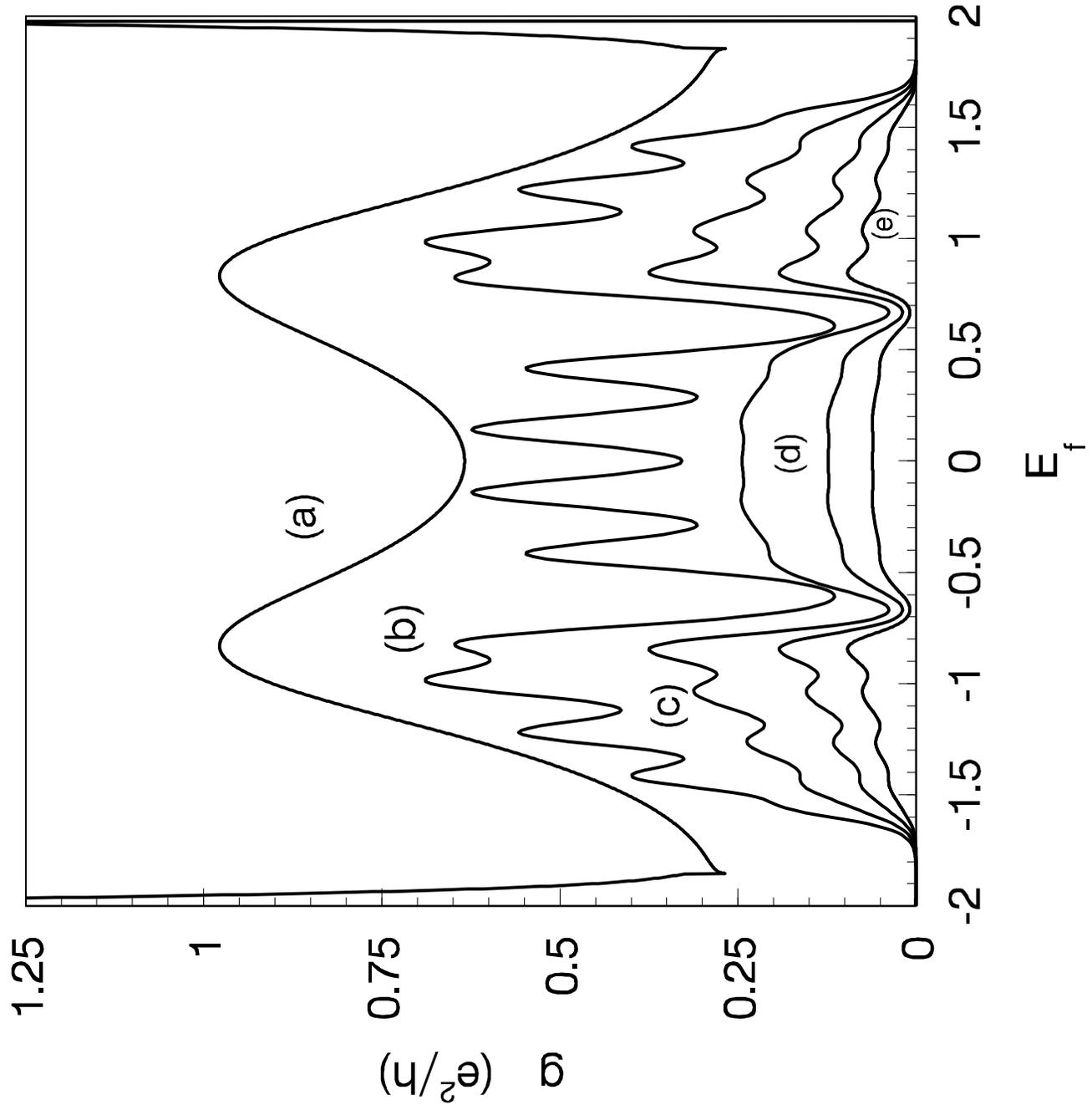
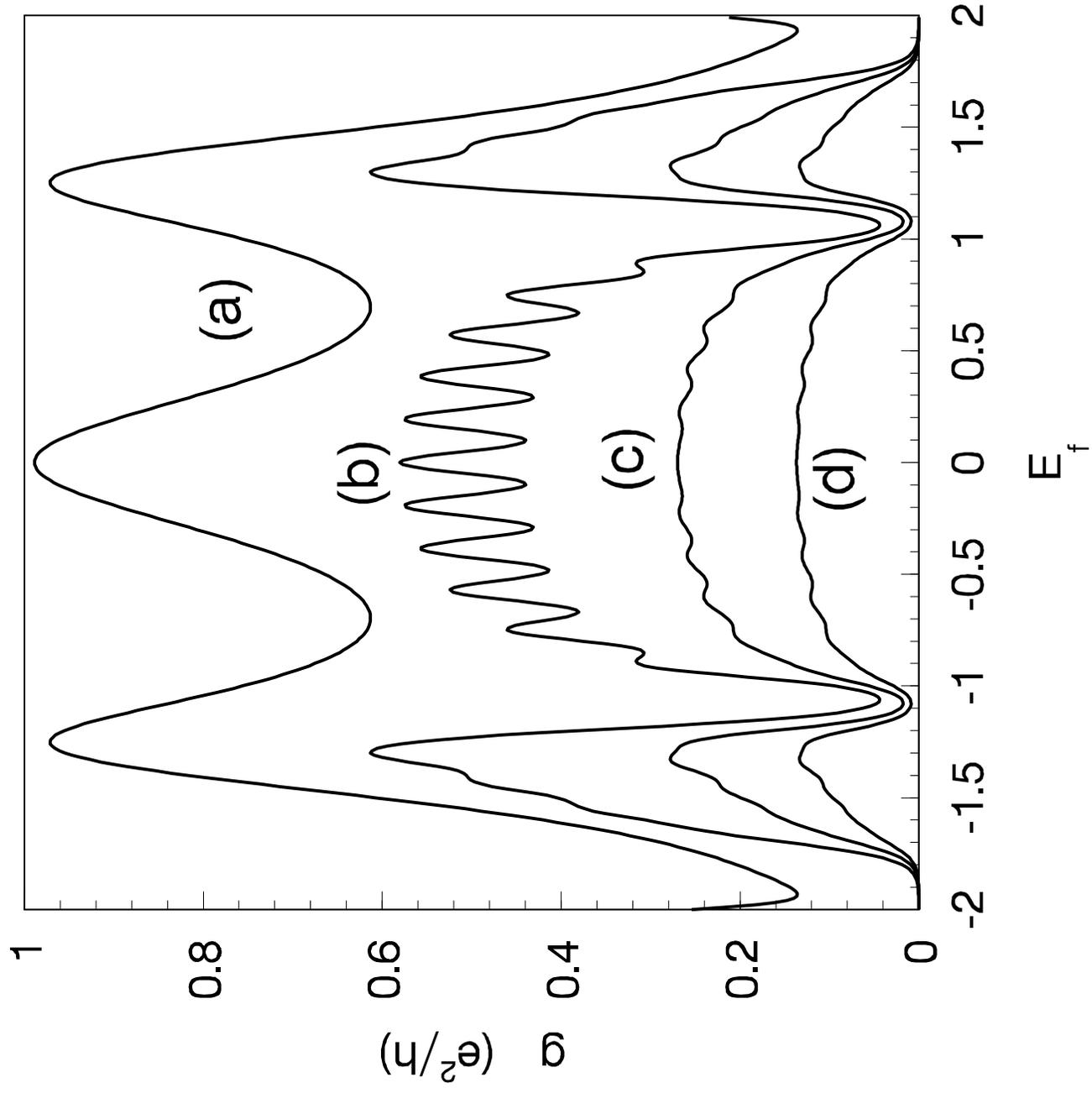
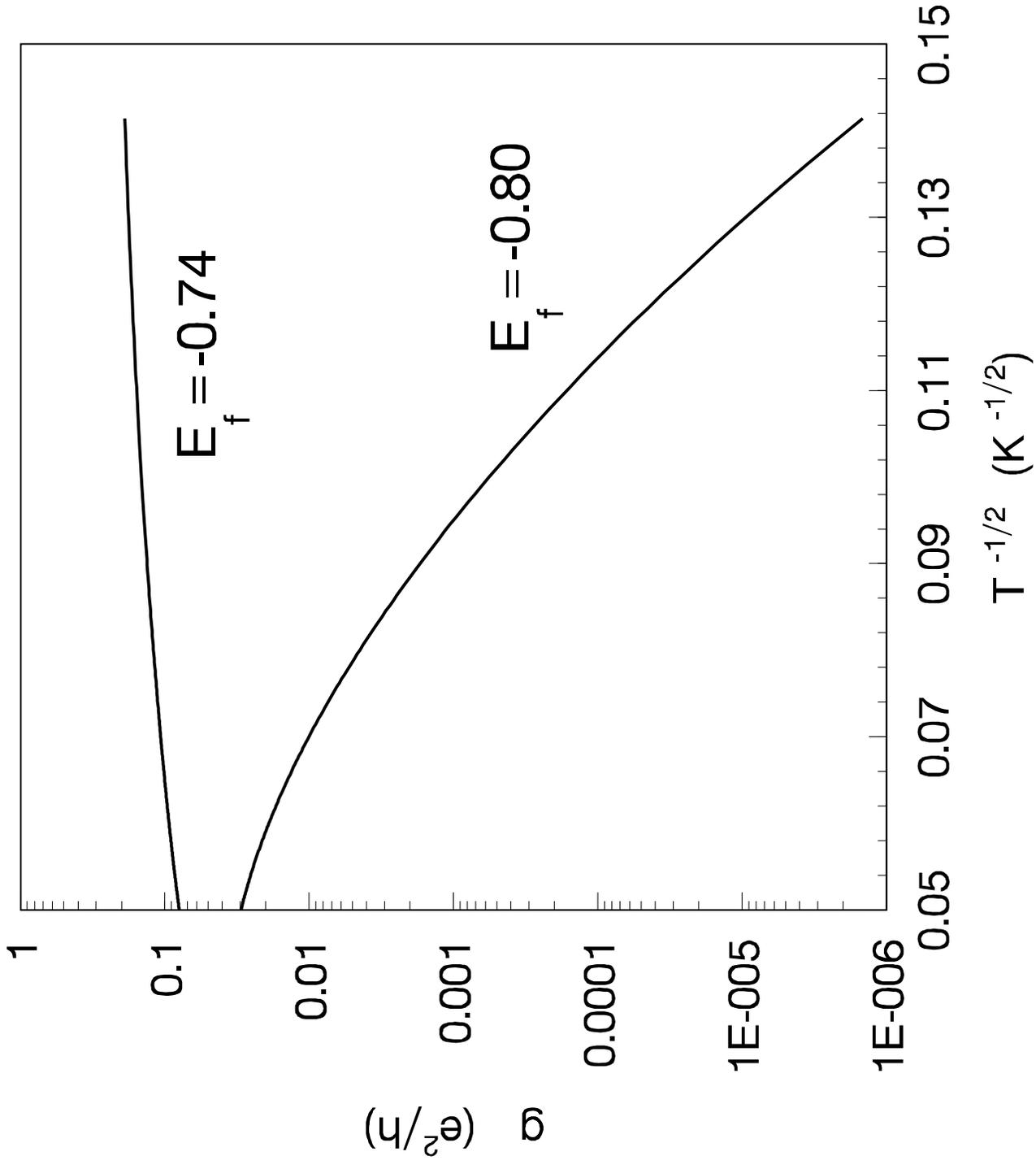
\end{document}